\pdfoutput=1

\documentclass[11pt]{article}

\usepackage[preprint]{acl}

\usepackage{times}
\usepackage{latexsym}

\usepackage[T1]{fontenc}

\usepackage[utf8]{inputenc}

\usepackage{microtype}

\usepackage{inconsolata}

\usepackage{graphicx}

\usepackage{booktabs}
\usepackage{multirow} 
\usepackage{adjustbox} 
\usepackage{array} 
\usepackage{amsmath}
\usepackage{tikzducks}
\usepackage{subcaption}
\usepackage{xspace}
\newcommand{\myname}{HeteRAG\xspace}

\title{HeteRAG: A Heterogeneous Retrieval-augmented Generation Framework with Decoupled Knowledge Representations }

\author{
 \textbf{Peiru Yang\textsuperscript{1}},
 \textbf{Xintian Li\textsuperscript{1}},
 \textbf{Zhiyang Hu\textsuperscript{2}},
 \textbf{Jiapeng Wang\textsuperscript{3}},
 \textbf{Jinhua Yin\textsuperscript{1}},
 \textbf{Huili Wang\textsuperscript{1}},
 \\
 \textbf{Lizhi He\textsuperscript{4}},
 \textbf{Shuai Yang\textsuperscript{4}},
 \textbf{Shangguang Wang\textsuperscript{3}},
 \textbf{Yongfeng Huang\textsuperscript{1}},
 \textbf{Tao Qi\textsuperscript{3,*}}
\\
\\
    \textsuperscript{1}Tsinghua University,
    \textsuperscript{2}Xinjiang University,
    \textsuperscript{3}Beijing University of Posts and Telecommunications, \\
    \textsuperscript{4}JD Health International Inc. 
\\
}

\begin{document}
\maketitle
\begin{abstract}
Retrieval-augmented generation (RAG) methods can enhance the performance of LLMs by incorporating retrieved knowledge chunks into the generation process. 
In general, the retrieval and generation steps usually have different requirements for these knowledge chunks.
The retrieval step benefits from comprehensive information to improve retrieval accuracy, whereas excessively long chunks may introduce redundant contextual information, thereby diminishing both the effectiveness and efficiency of the generation process.
However, existing RAG methods typically employ identical representations of knowledge chunks for both retrieval and generation, resulting in suboptimal performance.
In this paper, we propose a heterogeneous RAG framework (\myname) that decouples the representations of knowledge chunks for retrieval and generation, thereby enhancing the LLMs in both effectiveness and efficiency.
Specifically, we utilize short chunks to represent knowledge to adapt the generation step and utilize the corresponding chunk with its contextual information from multi-granular views to enhance retrieval accuracy.
We further introduce an adaptive prompt tuning method for the retrieval model to adapt the heterogeneous retrieval augmented generation process.
Extensive experiments demonstrate that \myname achieves significant improvements compared to baselines.


\end{abstract}

\begin{figure}[ht]
\centering
\includegraphics[width=\linewidth]{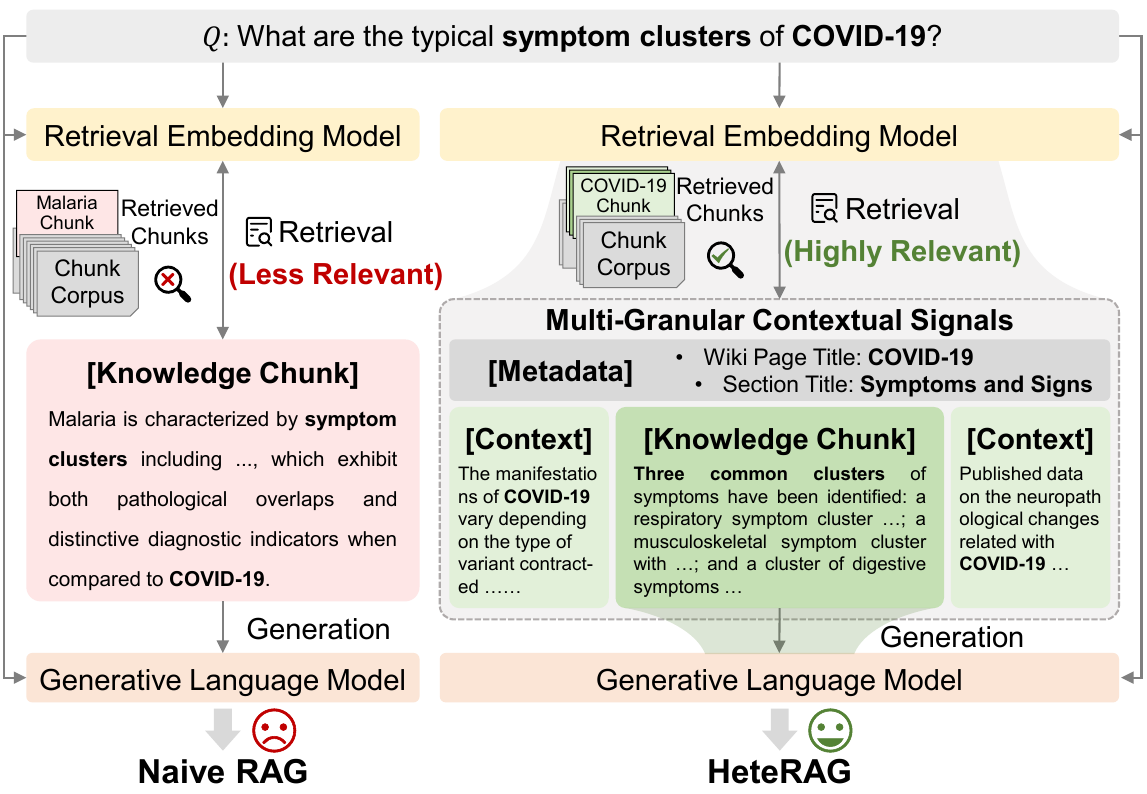} 
\caption{
Naive RAG suffers retrieval inaccuracy due to identical chunk representations for retrieval/generation.
The decoupled architecture of \myname addresses this via contextual signal- and metadata-enhanced retrieval.
}
\label{fig:intro} 
\end{figure}

\section{Introduction}
Retrieval Augmented Generation (RAG) technology is a powerful technique for building capable and reliable AI systems \cite{lewis2020retrieval}.
By incorporating external knowledge chunks into LLMs' generation process, RAG enables more accurate responses and effectively mitigates the occurrence of hallucinations.
RAG systems first segment the knowledge corpus into limited-size chunks, then retrieve relevant chunks by calculating the similarity between the user query and chunk encoded representations using the retrieval model.
These retrieved chunks are subsequently incorporated into the prompt for the LLM, allowing it to generate contextually informed responses.

%
In typical RAG architectures, the retrieval and generation phases demonstrate distinct requirements regarding knowledge chunk granularity.
As interactive objects of the retriever, knowledge chunks are required to accurately match user queries to help the retrieval model find the most relevant information.
Therefore, the retrieval step requires semantically complete information to ensure retrieval accuracy.
Conversely, excessively long chunks may introduce redundant or irrelevant information.
This may potentially induce hallucinations in LLMs \cite{huang2023survey}, thereby compromising the efficacy and efficiency of the generation process.
Hence, knowledge chunks are expected to provide the most precise information to answer the user’s questions.
However, most existing RAG methods employ identical representations of knowledge chunks for both retrieval and generation, and thus face challenges in jointly optimizing the performance of both stages caused by the identical granularity of knowledge chunk representation.

To address this problem, we propose \myname, a heterogeneous RAG framework that decouples the representations of knowledge chunks for retrieval and generation stages.
As shown in Fig~\ref{fig:intro}, we employ a context-enriched modeling strategy at retrieval side to integrate both multi-granular contextual signals and global structured metadata, enhancing the retrieval accuracy.
Meanwhile, we utilize standalone knowledge chunks for the generation process, enabling LLMs to generate with high efficiency and precision.
This architecture facilitates joint optimization of both stages.
Building on this, we further propose an adaptive prompt tuning strategy that enables the retrieval model to dynamically align with our context-enriched modeling strategy. 
It facilitates the specialization of off-the-shelf embedding models, allowing them to effectively handle diverse, structurally complex real-world knowledge corpus.
We conduct extensive experiments on retrieval tasks and end-to-end RAG pipelines to evaluate the effectiveness of \myname.
Experimental results demonstrate that \myname achieves significant improvements compared to baselines.
The consistent gains in retrieval and QA accuracy confirm \myname effectively resolves the two-stage optimization conflict, thereby enhancing the real-world applicability of RAG.
Our codes are available at: \url{https://anonymous.4open.science/r/HeteRAG/}. Our contributions can be summarized as follows: 
\begin{itemize}
\item[$\bullet$]We introduce a novel heterogeneous RAG framework that decouples knowledge representations for retrieval and generation step.
\item[$\bullet$]We design a prompt tuning strategy that adaptively aligns pre-trained models with the heterogeneous RAG process.
\item[$\bullet$] Extensive experiments on 3 knowledge bases, 5 datasets, 4 retrieval model, and 3 foundation models demonstrate that HeteRAG effectively outperforms baseline RAG methods.
\end{itemize}

\begin{figure*}[ht]
    \centering
    \includegraphics[width=\textwidth]{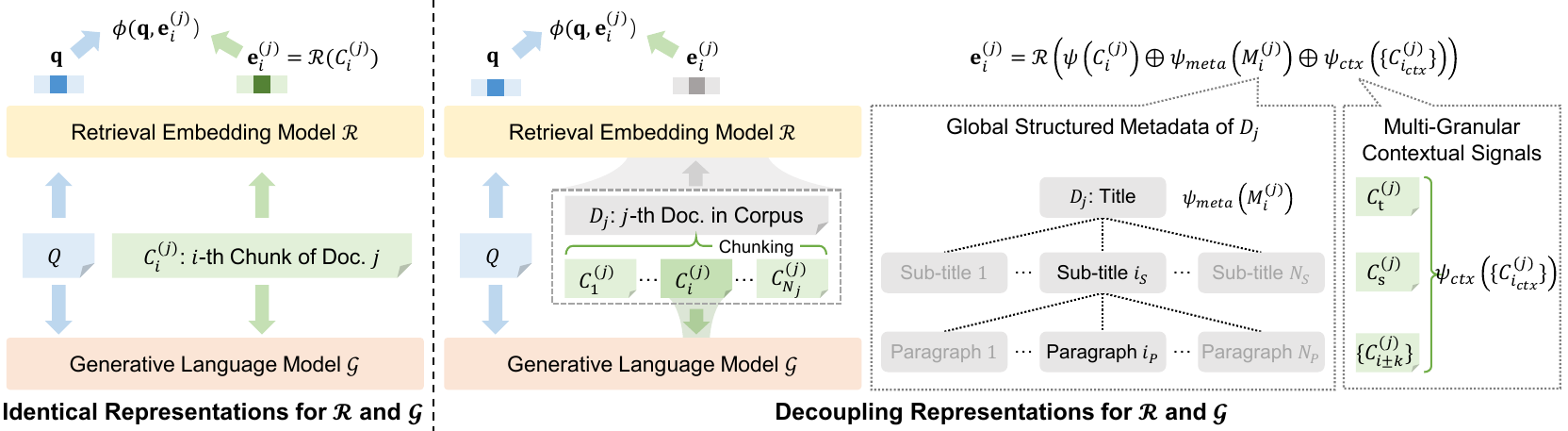} 
    \caption{
    The overall framework of \myname. 
    The left shows naive RAG using identical representations of knowledge chunks for retrieval and generation. 
    The right depicts \myname's framework: retrieval incorporates global metadata and multi-granular context, while generation maintains standalone chunk usage.
    }
    \label{fig:method}
\end{figure*}

\section{Related Works}
\subsection{Retrieval Models}
Retrieval models aim to retrieve relevant information from a corpus based on queries. 
Modern approaches predominantly employ transformer-based pre-trained embedding models for dense retrieval, a paradigm that learns latent space representations of queries and chunks through neural encoding. 

Recent progress features several impactful embedding models that demonstrate state-of-the-art performance across various benchmarks. 
The E5 family \cite{wang2022text} train text embeddings in a contrastive manner using weak supervision from a large-scale text pair dataset.
Jina Embeddings \cite{gunther2023jina} focus on long text input and extend token limits, effectively handling long documents without the need for truncation or paragraph splitting.
BGE embedding family \cite{xiao2024c, chen2024bge} is a versatile embedding model trained through multi-stages that exhibits highly competitive performance in multilingual and cross-lingual retrieval tasks.
These versatile embedding models are capable of uniformly supporting a variety of tasks, providing support for multiple applications, including RAG.
Note that our work is orthogonal to these embedding models; it can be implemented in any embedding model to enhance their performance in retrieval tasks.

\subsection{Retrieval Augmented Generation}
Since the RAG framework was first proposed \cite{lewis2020retrieval, guu2020retrieval}, it has become an important supporting technology in the real-world applications of LLMs.
By providing reliable and up-to-date external knowledge to LLMs, RAG effectively enhances their generation performance.
In recent years, many works have improved the retrieval stage of RAG through various optimization methods.
\citet{yu2023augmentation} introduce an augmentation-adapted retriever which is trained to learn unseen LLMs' preferences from a known source language model.
\citet{shi2024replug} append retrieved documents to the input of a frozen language model, differentiating itself from previous methods that train language models to adapt to retrievers.
Overall, these works still employ identical representations of knowledge chunks for both retrieval and generation stages.

Some works have decoupled retrieval and generation representations to a certain extent.
For example, late chunking method \cite{gunther2024late} utilizes long context embedding models to first embed all tokens before applying chunking, resulting in chunk embeddings that preserve full contextual information and improve performance on retrieval tasks.
However, the effectiveness of late chunking is limited when using conventional embedding models or in scenarios involving extremely long documents.
\citet{raina2024question} introduce a zero-shot adaptation of dense retrieval by decomposing chunks into atomic statements and generating synthetic questions for improved chunk recall.
\citet{chen2024hiqa} propose a multi-document QA framework with cascading metadata integration and multi-route retrieval for multi-document environments.
\citet{anthropic2024} present a method for generating contextualized chunk embeddings by using a large language model (LLM) to augment chunk text with relevant context from the entire document before embedding.
These works might face latency caused by the generation of KG or summaries by large models, which limits their effectiveness in online settings and with larger corpora.

\section{Methods}
In this section, we first give a problem formulation of retrieval and generation process of RAG. 
Then we elaborate on \myname framework in detail.

\subsection{Problem Formulation} 
Given a document corpus $\{ D_1, ..., D_M \}$ and user query $Q$, an RAG system operates through three coordinated phases: document chunking, dense vector retrieval, and conditional generation. 
A chunking strategy is used to first decompose each document $D_j$ into text chunks through a chunking strategy $\mathcal{C}$:
$ \{ C^{(j)}_1, ..., C^{(j)}_{N_j} \} = \mathcal{C}(D_j) \quad \forall j \in \{1,...,M\}$.
$C^{(j)}_i$ denotes the $i$-th chunk from document $D_j$, resulting in a global chunk collection $\bigcup_{j=1}^M \{ C^{(j)}_1, ..., C^{(j)}_{N_j} \}$. 
Then the retriever $\mathcal{R}$ encodes both the query and all chunks into a shared embedding $d$-dimensional vector space:
\begin{equation}
\mathbf{q} = \mathcal{R}(Q), \quad \mathbf{e}^{(j)}_i = \mathcal{R}(C^{(j)}_i)
\end{equation}
The system computes pairwise similarity scores $\phi(\mathbf{q}, \mathbf{e}^{(j)}_i)$ between the query embedding and chunk embeddings, typically implemented as cosine similarity. 
The top-$k$ most relevant chunks are passed to LLM $\mathcal{G}$ to generate the final response.

\subsection{Knolwedge Representation Decoupling} 
Our \myname framework addresses the representation dilemma by decoupling the representations of knowledge chunks for retrieval and generation. 
As illustrated in Fig~\ref{fig:method}, the architecture establishes dual pathways for retrieval-oriented and generation-oriented knowledge chunks, enabling specialized optimization for each stage.
After chunking the document corpus $\{D_j\}_{j=1}^M$ into the global chunk collection $\bigcup_{j=1}^M \{ C^{(j)}_1, ..., C^{(j)}_{N_j} \}$, we model the retrieval and generation stages separately.

The retrieval side aims to precisely align user queries with relevant documents, necessitating comprehensive information from the retrieval side to sufficiently model and compute semantic similarity. 
RAG systems across different tasks and domains typically operate on heterogeneous corpora with distinct structural characteristics. 
For instance, corpora may exhibit hierarchical tree structures (e.g., Wikipedia articles with nested sections), linear sequences (e.g., news articles with temporal dependencies), or graph-based organizations (e.g., knowledge bases with entity-relation networks). 
To effectively leverage such diversity, multi-granular information integration at the retrieval side becomes crucial – this typically encompasses raw knowledge chunks, multi-granular contextual signals, and global structured metadata, each contributing complementary perspectives for robust retrieval.
For a knowledge chunk $C^{(j)}_i$ in document $D_j$, We formulate the modeling procedure at the retrieval side as follows:
\begin{equation}
\label{eq:retrieval}
    \scalebox{0.85}{$\displaystyle
    \mathbf{e}_i^{(j)} = \mathcal{R}\left[\psi(C_i^{(j)}) \oplus \psi_{\text{ctx}}(\{C_{i_{ctx}}^{(j)}\}) \oplus \psi_{\text{meta}}(M_i^{(j)})\right]
    $}
\end{equation}
where \( M_i^{(j)} \) represents the global metadata of \( C_i^{(j)} \) in \( D_j \), including but not limited to subject, abstract, document title, section title, subsection title, related keywords, etc.
And \( \{ C_{i_{ctx}}^{(j)} \} = \{C_t^{(j)}, C_s^{(j)}, \{C_{i\pm k}^{(j)} \} \}\) represents the multi-granular contextual signals, which can provide the retrieval model with contextual information at different levels.
$\psi(\cdot)$, $\psi_{\text{ctx}}(\cdot)$, and $\psi_{\text{meta}}(\cdot)$ are the semantic encoders for different components, and $\oplus$ denotes the fusion operation.

The generation side aims to keep the representation of the knowledge chunk as concise as possible for the sake of efficiency and precision, avoiding redundant or unnecessary information.
Therefore, in contrast to the retrieval side, we only provide $C^{(j)}_i$ itself to the generative model on the generation side to maintain task-specific precision:
\begin{equation}
    \text{Ans}=\mathcal{G}\left(T(Q, C_i^{(j)})\right) 
\end{equation}
Where $T(\cdot, \cdot)$ refers to the prompt template accustomed to the generative language model $\mathcal{G}$.
In this way, the representations are decoupled between retrieval and generation.


\subsection{Adaptive Prompt tuning Strategy}
\label{subsec:finetune}
In many cases where RAG systems are applied to specific domains, the retrieval embedding model is fine-tuned to adapt to the corresponding domain.
To specialize the retrieval model for heterogeneous document structures, we introduce an adaptive fine-tune strategy.

Prompt tuning \cite{lester2021power} has been widely adopted in various fields, as it uses task-specific instructions to improve performance on targeted tasks.
To enable the retrieval model to better leverage contextual signals and structured metadata, as well as to adapt to the characteristics of different corpora, we propose a fine-tuning strategy based on prompt tuning.
Specifically, we prepend instructions to different information units of a certain chunk. 
For a chunk $C_i^{(j)}$ with contextual signals $\{C_{i\pm k}^{(j)}\}$ and global metadata $M_i^{(j)}$, we formulate the instruction input as: 
\begin{equation}
\tilde{C}_h = [\text{INST}_h] \oplus C
\end{equation}  
where $[\text{INST}_h]$ denotes the instruction embedding specific to hierarchy level $h$, implemented as soft prompts through continuous token vectors. The retrieval model $\mathcal{R}$ then encodes both the original query $Q$ and prompted chunks $\{\tilde{C}_h\}$ into an adaptive embedding space:  
\begin{equation}
\mathbf{q} = \mathcal{R}(Q), \quad \tilde{\mathbf{e}}_h = \mathcal{R}(\tilde{C}_h)
\end{equation}

Following the conventional paradigm of contrastive learning, we construct positive and negative samples.
Given a user query $Q$, the positive pair \((Q, C^+)\) is directly derived from human-annotated relevance data. 
For negative pairs \((Q, C^-)\), both in-batch negatives \( \{C_j^-\}_{j\neq i} \)and random negatives \( C_{\text{rand}}^- \) are employed for training.

Similarity between \( Q \) and \( C \) is measured by scaled cosine similarity:  
\begin{equation}
\phi(Q,C) = \frac{\mathbf{q}^\top \tilde{\mathbf{e}}_h}{\|\mathbf{q}\|\|\tilde{\mathbf{e}}_h\|} \cdot \tau^{-1}
\end{equation} 
where \( \tau \) denotes the temperature hyperparameter controlling the softness of the similarity distribution. 
The model is trained using an InfoNCE loss \cite{oord2018representation}:  
\begin{equation}
    \scalebox{0.8}{$\displaystyle \mathcal{L} = -\frac{1}{N}\sum_{i=1}^N \log \frac{e^{s(Q_i,C_i^+)}}{\sum_{j=1}^N e^{s(Q_i,C_j^+)} + \sum_{k=1}^K e^{s(Q_i,C_k^-)}}$}
\end{equation}


\begin{table*}[t]
      \centering
       \begin{adjustbox}{width=\textwidth}
    \begin{tabular}{c|c|c|cc|cc|cc|cc}
    \toprule
    \multirow{2}[4]{*}{Dataset} & \multicolumn{1}{c|}{\multirow{2}[4]{*}{Emb\newline{}Model}} & \multirow{2}[4]{*}{Method} & \multicolumn{2}{c|}{chunk size=16} & \multicolumn{2}{c|}{chunk size=32} & \multicolumn{2}{c|}{chunk size=64} & \multicolumn{2}{c}{chunk size=128} \\
\cmidrule{4-11}          &       &       & nDCG@1 & nDCG@10 & nDCG@1 & nDCG@10 & nDCG@1 & nDCG@10 & nDCG@1 & nDCG@10 \\
    \midrule
    \midrule
    \multirow{12}[8]{*}{\rotatebox{90}{SciFact}} & \multicolumn{1}{c|}{\multirow{3}[2]{*}{Jina}} & Naive & 45.33\% & 58.74\% & 53.33\% & 64.23\% & 56.00\% & 66.29\% & 53.00\% & 64.79\% \\
          &       & Late  & 55.00\% & 66.63\% & 55.33\% & 66.86\% & 54.00\% & 66.05\% & 54.67\% & 66.12\% \\
          &       & \myname  & \textbf{58.67\%} & \textbf{68.90\%} & \textbf{57.33\%} & \textbf{68.51\%} & \textbf{57.00\%} & \textbf{67.83\%} & \textbf{56.00\%} &  \textbf{67.50}\% \\
\cmidrule{2-11}          & \multicolumn{1}{c|}{\multirow{3}[2]{*}{BGE}} & Naive & 53.67\% & 66.76\% & 57.33\% & 69.68\% & 59.00\% & 70.87\% & \textbf{61.33\%} & 73.09\% \\
          &       & Late  & 60.00\% & 72.10\% & 59.33\% & 71.91\% & 59.00\% & 71.70\% & 59.33\% & 71.94\% \\
          &       & \myname  & \textbf{63.00\%} & \textbf{74.54\%} & \textbf{64.33\%} & \textbf{75.89\%} & \textbf{64.00\%} & \textbf{75.54\%} & 60.33\% & \textbf{73.49\%} \\
\cmidrule{2-11}          & \multicolumn{1}{c|}{\multirow{3}[2]{*}{E5}} & Naive & 44.00\% & 58.53\% & 52.33\% & 64.03\% & 51.33\% & 63.75\% & 47.67\% & 58.90\% \\
          &       & Late  & 53.00\% & 66.79\% & 53.67\% & 66.77\% & 53.00\% & 66.79\% & 52.67\% & 66.56\% \\
          &       & \myname  & \textbf{60.33\%} & \textbf{71.74\%} & \textbf{60.00\%} & \textbf{71.04\%} & \textbf{58.67\%} & \textbf{70.16\%} & \textbf{52.67\%} & \textbf{66.82\%} \\
\cmidrule{2-11}          & \multirow{3}[2]{*}{MedEmb} & Naive & 50.33\% & 62.94\% & 56.00\% & 66.80\% & 56.00\% & 68.41\% & 58.33\% & 69.85\% \\
          &       & Late  & 57.33\% & 68.96\% & 57.33\% & 69.03\% & 57.67\% & 68.82\% & 57.00\% & 68.47\% \\
          &       & \myname  & \textbf{66.59\%} & \textbf{71.18\%} & \textbf{62.00\%} & \textbf{72.31\%} & \textbf{61.33\%} & \textbf{72.15\%} & \textbf{60.00\%} & \textbf{71.70\%} \\
    \midrule
    \midrule
    \multicolumn{1}{c|}{\multirow{12}[8]{*}{\rotatebox{90}{NF-\newline{}Corpus}}} & \multicolumn{1}{c|}{\multirow{3}[2]{*}{Jina}} & Naive & 32.51\% & 25.24\% & 29.10\% & 24.00\% & 31.27\% & 24.40\% & 30.96\% & 24.01\% \\
          &       & Late  & 40.56\% & 31.20\% & 41.33\% & 30.84\% & 39.94\% & 30.73\% & 39.47\% & \textbf{30.33\%} \\
          &       & \myname  & \textbf{41.95\%} & \textbf{31.98\%} & \textbf{43.65\%} & \textbf{32.07\%} & \textbf{40.25\%} & \textbf{30.92\%} & \textbf{39.78\%} & 29.81\% \\
\cmidrule{2-11}          & \multicolumn{1}{c|}{\multirow{3}[2]{*}{BGE}} & Naive & 41.95\% & 33.49\% & 43.34\% & 34.68\% & 44.12\% & 35.16\% & 41.64\% & 35.47\% \\
          &       & Late  & 46.29\% & 36.68\% & 45.98\% & 36.60\% & 45.67\% & 36.46\% & 44.89\% & 36.41\% \\
          &       & \myname  & \textbf{48.45\%} & \textbf{37.65\%} & \textbf{49.38\%} & \textbf{37.66\%} & \textbf{47.52\%} & \textbf{37.65\%} & \textbf{46.75\%} & \textbf{37.01\%} \\
\cmidrule{2-11}          & \multicolumn{1}{c|}{\multirow{3}[2]{*}{E5}} & Naive & 39.63\% & 30.72\% & 32.97\% & 29.42\% & 32.51\% & 28.50\% & 32.35\% & 26.08\% \\
          &       & Late  & 39.01\% & 31.15\% & 38.70\% & 30.93\% & 37.31\% & 30.69\% & 35.91\% & 30.17\% \\
          &       & \myname  & \textbf{43.81\%} & \textbf{36.07\%} & \textbf{44.58\%} & \textbf{35.49\%} & \textbf{44.89\%} & \textbf{35.84\%} & \textbf{43.34\%} & \textbf{34.86\%} \\
\cmidrule{2-11}          & \multirow{3}[2]{*}{MedEmb} & Naive & 44.12\% & 33.44\% & 43.50\% & 33.25\% & 43.96\% & 33.03\% & 41.33\% & 32.55\% \\
          &       & Late  & 43.19\% & 34.51\% & 42.42\% & 34.39\% & 41.49\% & 34.26\% & 41.02\% & 33.82\% \\
          &       & \myname  & \textbf{46.59\%} & \textbf{35.62\%} & \textbf{47.37\%} & \textbf{35.83\%} & \textbf{43.96\%} & \textbf{35.18\%} & \textbf{43.65\%} & \textbf{34.94\%} \\
    \midrule
    \midrule
    \multicolumn{1}{c|}{\multirow{12}[8]{*}{\rotatebox{90}{Trec-\newline{}COVID}}} & \multicolumn{1}{c|}{\multirow{3}[2]{*}{Jina}} & Naive & 56.00\% & 51.82\% & 55.00\% & 52.82\% & 58.00\% & 60.55\% & 65.00\% & 64.16\% \\
          &       & Late  & \textbf{74.00\%} & 66.91\% & 65.00\% & 66.23\% & 73.00\% & 67.66\% & 77.00\% & 67.12\% \\
          &       & \myname  & 73.00\% & \textbf{69.26\%} & \textbf{72.00\%} & \textbf{69.79\%} & \textbf{77.00\%} & \textbf{71.77\%} & \textbf{81.00\%} & \textbf{70.31\%} \\
\cmidrule{2-11}          & \multicolumn{1}{c|}{\multirow{3}[2]{*}{BGE}} & Naive & 68.00\% & 62.60\% & 66.00\% & 62.37\% & 65.00\% & 65.06\% & 66.00\% & 67.07\% \\
          &       & Late  & 70.00\% & 64.93\% & 67.00\% & 46.30\% & 73.00\% & 70.01\% & 69.00\% & 67.62\% \\
          &       & \myname  & \textbf{78.00\%} & \textbf{76.60\%} & \textbf{86.00\%} & \textbf{75.33\%} & \textbf{87.00\%} & \textbf{77.30\%} & \textbf{82.00\%} & \textbf{75.97\%} \\
\cmidrule{2-11}          & \multicolumn{1}{c|}{\multirow{3}[2]{*}{E5}} & Naive & 67.00\% & 57.03\% & 63.00\% & 54.75\% & 58.00\% & 51.62\% & 55.00\% & 51.66\% \\
          &       & Late  & 57.00\% & 46.50\% & 61.00\% & 34.16\% & \textbf{59.00\%} & 49.70\% & \textbf{60.00\%} & 51.28\% \\
          &       & \myname  & \textbf{69.00\%} & \textbf{63.23\%} & \textbf{68.00\%} & \textbf{61.99\%} & 56.00\% & \textbf{57.32\%} & 55.00\% & \textbf{54.96\%} \\
\cmidrule{2-11}          & \multirow{3}[2]{*}{MedEmb} & Naive & 57.00\% & 60.77\% & 67.00\% & 65.51\% & 67.00\% & 67.03\% & 75.00\% & 72.14\% \\
          &       & Late  & 73.00\% & 66.19\% & 75.00\% & 46.41\% & 72.00\% & 65.37\% & 76.00\% & 67.32\% \\
          &       & \myname  & \textbf{79.00\%} & \textbf{74.58\%} & \textbf{81.00\%} & \textbf{76.17\%} & \textbf{87.00\%} & \textbf{79.47\%} & \textbf{83.00\%} & \textbf{78.81\%} \\
    \bottomrule
    \end{tabular}%
\end{adjustbox}
  \caption{
  Evaluation of different chunk representation methods on retrieval tasks. 
  \myname significantly improves retrieval accuracy in the majority of settings.
  }
\label{tab:retrieval}%
\end{table*}%

\begin{table*}[t]
  \centering
  \begin{adjustbox}{width=\textwidth}
    \begin{tabular}{c|c|c|cc|cc|cc|cc}
    \toprule
    \multirow{2}[4]{*}{Dataset} & \multirow{2}[4]{*}{Emb} & \multirow{2}[4]{*}{Method} & \multicolumn{2}{c|}{chunk size=16} & \multicolumn{2}{c|}{chunk size=32} & \multicolumn{2}{c|}{chunk size=64} & \multicolumn{2}{c}{chunk size=128} \\
\cmidrule{4-11}          &       &       & ndcg@1 & ndcg@10 & ndcg@1 & ndcg@10 & ndcg@1 & ndcg@10 & ndcg@1 & ndcg@10 \\
    \midrule
    \multirow{9}[6]{*}{SciFact} & \multirow{3}[2]{*}{Jina} & Naive & 47.33\% & 61.87\% & 52.67\% & 64.70\% & 51.00\% & 66.06\% & 51.33\% & 65.53\% \\
          &       & Late    & 55.67\% & \textbf{70.89\%} & 56.00\% & 70.70\% & 56.33\% & 70.33\% & 54.33\% & 69.76\% \\
          &       & \myname  & \textbf{56.67\%} & 70.21\% & \textbf{58.00\%} & \textbf{71.98\%} & \textbf{59.00\%} & \textbf{72.01\%} & \textbf{60.00\%} & \textbf{71.87\%} \\
\cmidrule{2-11}          & \multirow{3}[2]{*}{BGE} & Naive & 49.00\% & 65.15\% & 58.67\% & 70.78\% & 62.67\% & 73.91\% & 63.33\% & 74.93\% \\
          &       & Late    & 61.33\% & 73.40\% & 62.00\% & 73.67\% & 61.67\% & 73.45\% & 61.00\% & 73.22\% \\
          &       & \myname  & \textbf{64.67\%} & \textbf{77.11\%} & \textbf{65.33\%} & \textbf{77.34\%} & \textbf{65.00\%} & \textbf{77.59\%} & \textbf{65.00\%} & \textbf{77.46\%} \\
\cmidrule{2-11}          & \multirow{3}[2]{*}{e5} & Naive & 47.67\% & 63.05\% & 53.00\% & 68.01\% & 55.33\% & 70.40\% & 58.00\% & 71.66\% \\
          &       & Late    & 56.67\% & 70.11\% & 56.00\% & 69.65\% & 55.33\% & 69.41\% & 55.00\% & 69.12\% \\
          &       & \myname  & \textbf{59.33\%} & \textbf{73.06\%} & \textbf{61.67\%} & \textbf{74.61\%} & \textbf{63.00\%} & \textbf{74.54\%} & \textbf{62.67\%} & \textbf{74.85\%} \\
    \bottomrule
    \end{tabular}%
    \end{adjustbox}
  \caption{
  Evaluation of our proposed adaptive fine-tune strategy on retrieval tasks.
  While fine-tuning generally enhances retrieval task performance, \myname still achieves superior results compared to the fine-tuned baselines.
  }
  \label{tab:finetune}%
\end{table*}%

\section{Experiments and Analysis}
\subsection{Experimental Datasets and Settings}
We utilize three information retrieval datasets for evaluation in the BEIR benchmark \cite{thakur2021beir}.
SciFact \cite{wadden2020fact} provides expert-written scientific claims with evidence-annotated research abstracts for claim verification.
NF-Corpus \cite{boteva2016full} focuses on medical information retrieval, while Trev-COVID \cite{voorhees2021trec} specializes in COVID-19-related retrieval.
Three widely-used embedding models are employed: E5-base-v2 \cite{wang2022text}, BGE-base-en-v1.5 \cite{xiao2024c}, and Jina-embeddings-v2-small \cite{gunther2023jina}.
We also utilize a specialized embedding model MedEmbed-small-v0.1 \cite{balachandran2024medembed} for medical and clinical corpus.
Among them, Jina is a long text embedding model with a capacity of 8192 tokens, while both E5, BGE, and MedEmb are regular models with a capacity of 512 tokens.

We conducted our end-to-end RAG experiments on five widely-used datasets:
PopQA \cite{mallen2023not} is a curated question set from diverse online platforms.
NQ dataset \cite{kwiatkowski2019natural} is a collection of real user queries paired with Wikipedia passages.
SQuAD \cite{rajpurkar2018know} is a widely-used benchmark dataset for machine comprehension, consisting of questions on a set of Wikipedia articles.
TriviaQA \cite{joshi2017triviaqa} contains 95K trivia-based QA pairs, while HotpotQA \cite{yang2018hotpotqa} offers 113K Wikipedia QA pairs for multi-hop reasoning challenges.
We used three state-of-the-art open-source LLMs as generative models: Llama3-8b-Instruct \cite{dubey2024llama}, Mistral-8B-Instruct \cite{jiang2024mixtral}, and Gemma-9b-Instruct \cite{team2024gemma}.
The end-to-end RAG code implementation refers to \citet{FlashRAG}.

Next, we introduce the experimental settings.
For retrieval experiments, we use commonly used ranking metrics ndcg@1 and ndcg@10.
For the adaptive tuning process, we conducted fine-tuning using the training partition of the SciFact dataset, followed by performance evaluation on the designated test partition.
For generation experiments, we use commonly used metrics in QA systems, namely EM (Exact Match) and token-level F1.
The token-level F1 metric refers to the harmonic mean of token-level precision and recall, calculated by comparing shared tokens between the response and golden answer.
In the retrieval corpus, we choose the widely-used Wiki2018 corpus, which is compatible with the five QA datasets used in the experiment.
To streamline the experiments, we select the first 1,000 samples from the test or development set of all QA datasets.
For vector database index building, we employ the Faiss library \cite{douze2024faiss}.
All experiments were conducted on four RTX 5000 GPUs.

\begin{table*}[t]
  \centering
  \begin{adjustbox}{width=0.95\textwidth}
    \begin{tabular}{c|cc|ccc|ccc|ccc}
    \toprule
    \multirow{2}[4]{*}{Model} & \multicolumn{2}{c|}{\multirow{2}[4]{*}{Dataset}} & \multicolumn{3}{c|}{w/o RAG} & \multicolumn{3}{c|}{Naive RAG} & \multicolumn{3}{c}{\myname} \\
\cmidrule{4-12}          & \multicolumn{2}{c|}{} & EM    & F1    & Recall & EM    & F1    & Recall & EM    & F1    & Recall \\
    \midrule
    \multirow{5}[2]{*}{\rotatebox{90}{Llama3-8b}} & \multicolumn{2}{c|}{PopQA} & 18.70\% & 22.96\% & 25.80\% & 24.00\% & 39.75\% & 58.66\% & 32.70\% & 52.25\% & 76.19\% \\
          & \multicolumn{2}{c|}{HotpotQA} & 19.70\% & 28.03\% & 28.06\% & 21.70\% & 30.56\% & 32.53\% & 30.80\% & 42.48\% & 43.32\% \\
          & \multicolumn{2}{c|}{TriviaQA} & 51.60\% & 58.94\% & 60.47\% & 52.40\% & 61.04\% & 63.65\% & 58.70\% & 68.56\% & 71.87\% \\
          & \multicolumn{2}{c|}{Squad} & 20.40\% & 27.09\% & 28.48\% & 28.90\% & 36.49\% & 40.11\% & 32.60\% & 40.34\% & 44.17\% \\
          & \multicolumn{2}{c|}{NQ} & 22.40\% & 32.61\% & 37.45\% & 29.80\% & 40.25\% & 47.01\% & 36.10\% & 48.24\% & 57.46\% \\
    \midrule
    \multirow{5}[2]{*}{\rotatebox{90}{Mistral-8b}} & \multicolumn{2}{c|}{PopQA} & 20.10\% & 22.51\% & 22.69\% & 32.70\% & 45.77\% & 58.94\% & 46.20\% & 61.40\% & 76.04\% \\
          & \multicolumn{2}{c|}{HotpotQA} & 18.60\% & 26.63\% & 26.30\% & 26.80\% & 37.21\% & 36.96\% & 36.60\% & 47.99\% & 47.91\% \\
          & \multicolumn{2}{c|}{TriviaQA} & 47.30\% & 53.90\% & 54.67\% & 55.40\% & 63.51\% & 64.87\% & 61.30\% & 69.53\% & 71.48\% \\
          & \multicolumn{2}{c|}{Squad} & 15.50\% & 21.75\% & 22.78\% & 33.30\% & 40.40\% & 42.52\% & 37.20\% & 44.24\% & 46.35\% \\
          & \multicolumn{2}{c|}{NQ} & 17.00\% & 24.68\% & 27.91\% & 33.00\% & 42.38\% & 47.22\% & 40.20\% & 51.54\% & 56.69\% \\
    \midrule
    \multirow{5}[2]{*}{\rotatebox{90}{gemma-9b}} & \multicolumn{2}{c|}{PopQA} & 15.00\% & 16.20\% & 16.40\% & 38.60\% & 48.16\% & 58.58\% & 52.00\% & 63.27\% & 75.51\% \\
          & \multicolumn{2}{c|}{HotpotQA} & 16.70\% & 24.39\% & 23.85\% & 25.10\% & 33.74\% & 33.07\% & 34.90\% & 45.40\% & 44.67\% \\
          & \multicolumn{2}{c|}{TriviaQA} & 52.40\% & 58.01\% & 58.09\% & 58.10\% & 64.79\% & 65.41\% & 63.60\% & 71.31\% & 72.22\% \\
          & \multicolumn{2}{c|}{Squad} & 16.30\% & 21.23\% & 22.03\% & 34.50\% & 39.63\% & 40.67\% & 37.90\% & 43.36\% & 44.61\% \\
          & \multicolumn{2}{c|}{NQ} & 21.60\% & 31.02\% & 32.70\% & 33.20\% & 42.72\% & 45.71\% & 39.80\% & 50.34\% & 54.22\% \\
    \bottomrule
    \end{tabular}%
    \end{adjustbox}
      \caption{
  The performance evaluation of different methods on five datasets and three LLMs.
  Across all datasets and models, \myname demonstrates higher QA accuracy on all evaluation metrics.
  }
  \label{tab:generation}%
\end{table*}%

\subsection{Performance Evaluation}
We conduct comprehensive experiments to evaluate the effectiveness of \myname on the BeIR benchmark, comparing against two baseline retrieval methods: naive RAG and late chunking.
Late chunking method \cite{gunther2024late} embeds all tokens in a document before applying chunking with a long text embedding model, to preserve full contextual information and improve retrieval performance.
For the Jina model, since it is specifically designed for long texts, late chunking can be applied directly. For the other two models, a variant called long late chunking is used, which employs a sliding window approach to concatenate embeddings.
Table~\ref{tab:retrieval} presents the retrieval performance across three representative datasets (SciFact for scientific claims, nfCorpus for medical information, and TREC COVID for COVID-19-related articles) using three embedding models with distinct architectures: Jina-v2 (long-text optimized), E5-v2, and BGE-v1.5 (both standard-length models).

From the experimental results, we made the following observations:
First, \myname consistently outperforms baseline methods in almost all cases.
Our method achieves average improvements of 9.43\% (nDCG@1) and 7.76\% (nDCG@10) over naive RAG across all datasets and models, with particularly notable gains on TrecCOVID (+11.73\% nDCG@10).
While the absolute performance of all three embedding models varies due to their inherent capacity differences, \myname maintains stable relative advantages regardless of the backbone model, suggesting effective decoupling of knowledge chunk modeling strategy from fundamental capabilities of embedding model.
This may be because of the context-enriched strategy of \myname on the retrieval side successfully models more comprehensive and rich information, thereby increasing recall accuracy.
Second, the late chunking method shows better performance on long text embedding models (Jina-v2) compared to naive RAG; however, on regular embedding models (E5-v2 and BGE-v1.5), the performance of the late chunking method declines.
We attribute this to the mismatch between the full-document encoding of late chunking (which Jina-v2 natively supports) and the sequence length constraints of regular models.
Furthermore, the late chunking method only applies to embedding models that use mean pooling and performs poorly on CLS-pooling models.
In contrast, \myname achieves better model-agnostic robustness. 
Third, varying chunk sizes from 16 to 128 tokens cause fluctuations in the performance of naive RAG.
Overall, smaller chunk sizes lead to lower retrieval performance due to the reduced amount of information.
Late chunking is less affected by chunk size due to its global modeling characteristics.
\myname also demonstrates strong stability through its multi-granular retrieval side modeling.
In other words, \myname can effectively adapt to different chunking sizes and strategies corresponding to various corpora.
These findings collectively validate advantages of \myname in cross-domain generalization, model compatibility, and operational robustness for real-world retrieval scenarios.

\begin{figure*}[!h]
\centering
\includegraphics[width=0.97\textwidth]{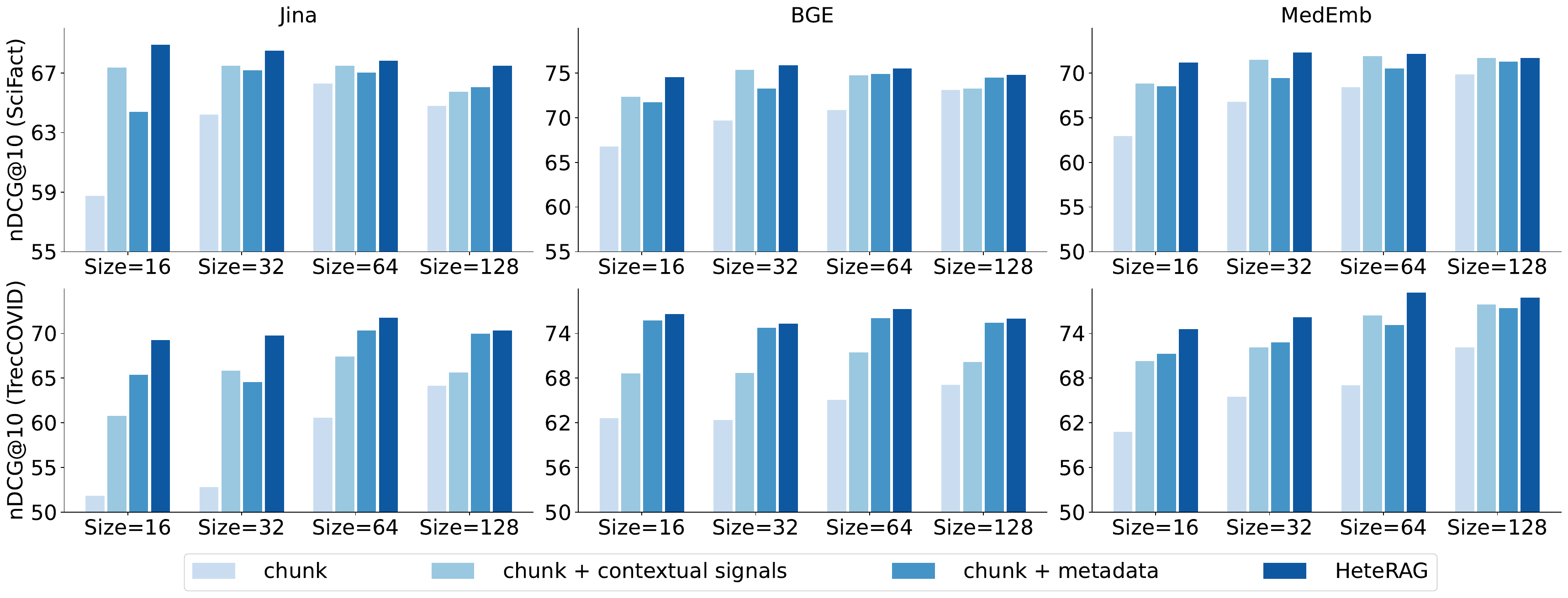} 
\caption{
Effect of contextual signals and structured metadata in \myname framework.
The ablation results show that both contribute significantly to the retrieval performance of \myname.
} 
\label{fig:abalation} 
\end{figure*}

\subsection{Evaluation on Adaptive Prompt Tuning}
As demonstrated in Table~\ref{tab:finetune}, the experimental results validate the effectiveness of the fine-tuning strategy described in Section~\ref{subsec:finetune} for \myname.
We implement contrastive learning-based fine-tuning for fair comparison on baseline methods following the same protocol.
Fine-tuned variants consistently outperform their non-fine-tuned counterparts across all datasets.
When trained with identical optimization steps, \myname achieves superior performance compared to fine-tuned baseline methods, confirming the benefits of our proposed fine-tuning strategy.
These findings demonstrate that \myname maintains compatibility with standard embedding model fine-tuning strategies, exhibiting strong adaptation capabilities.

\begin{figure*}[t]
  \centering
  \begin{subfigure}{0.4\textwidth}
    \includegraphics[width=\linewidth]{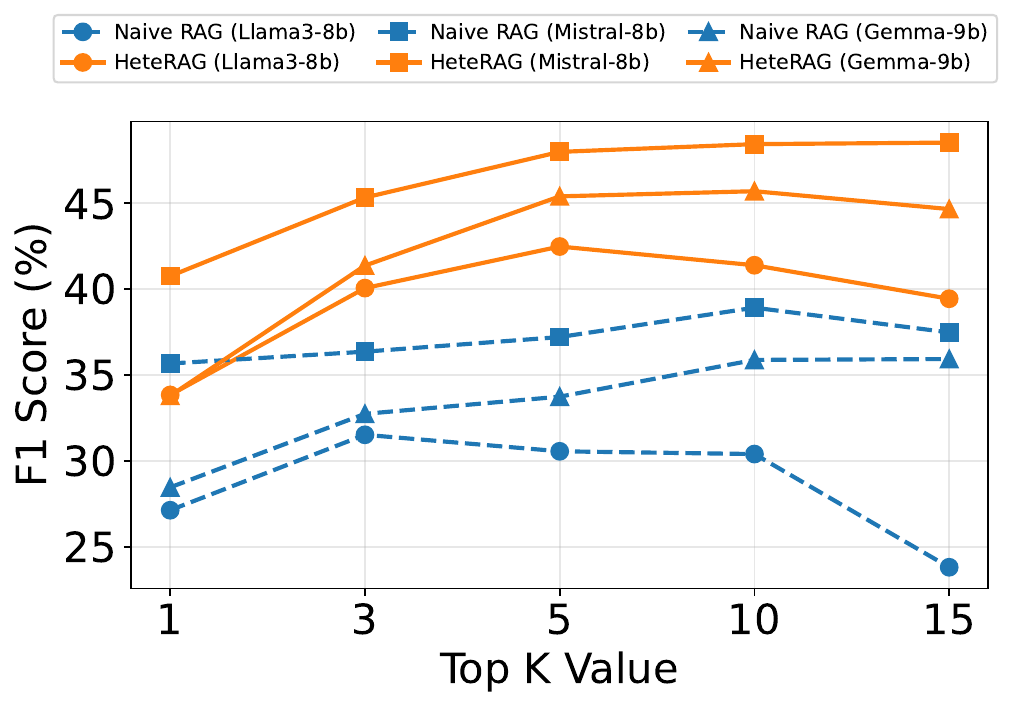}
  \end{subfigure}
  \begin{subfigure}{0.59\textwidth}
    \includegraphics[width=\linewidth]{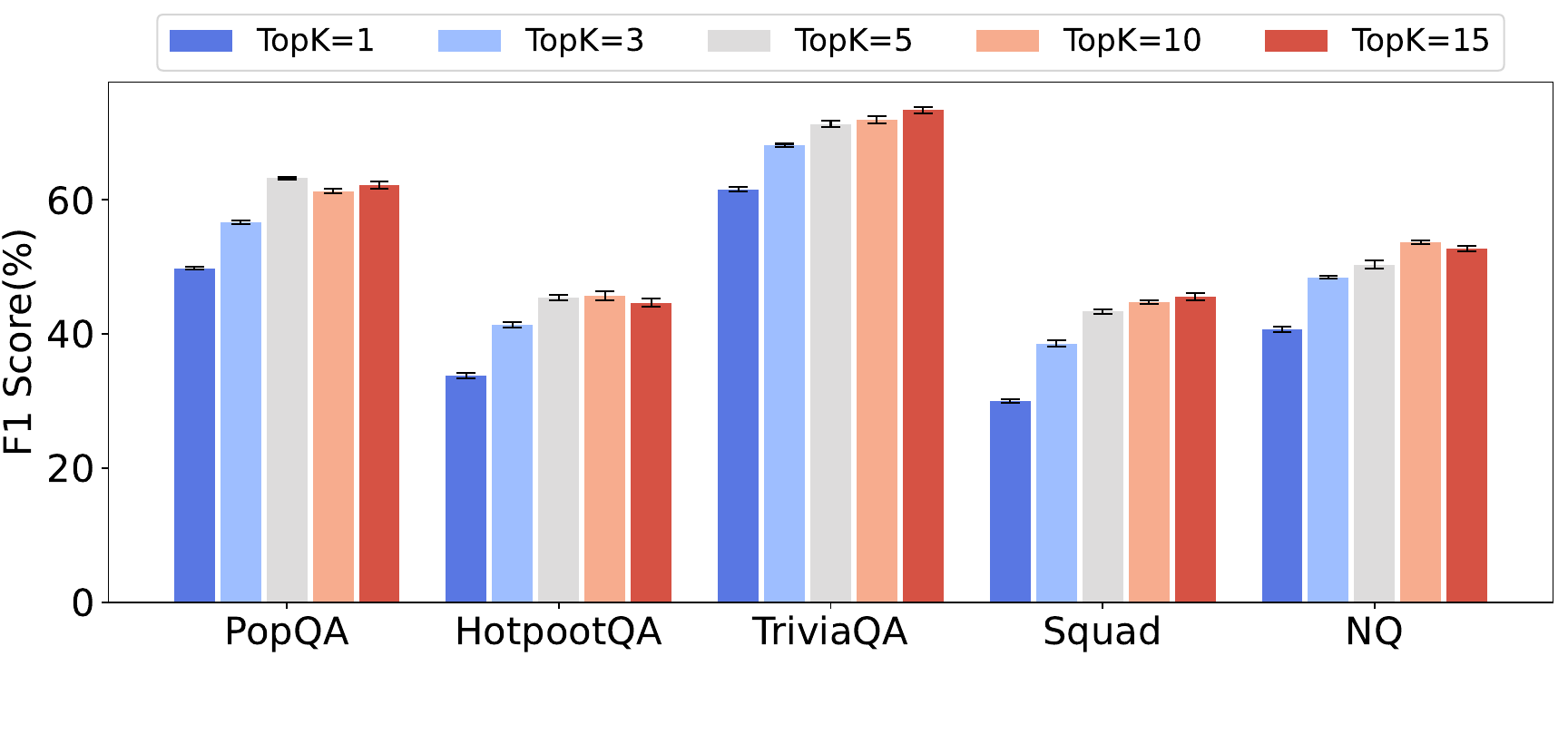}
  \end{subfigure}
  \caption{
  The RAG results under varying retrieval numbers (top-$k$).
  The left side shows the results of three LLMs on the HotpotQA dataset as they vary with top-$k$, using both naive RAG and \myname.
  The right side displays the performance variation of \myname across five different datasets under various top-$k$ settings.
  }
  \label{fig:topk}
\end{figure*}

\subsection{End-to-End RAG Performance}
The experimental results of our end-to-end RAG framework, as shown in Table~\ref{tab:generation}, demonstrate consistent performance improvements across three generative language models (Llama3-8b-Instruct, Mistral-8B-Instruct, and Gemma-9b-Instruct) and five benchmark datasets (NQ, PopQA, SQuAD, TriviaQA, and HotpotQA).
The table presents the results of retrieval top-5 knowledge chunks from the Wiki corpus.
\myname significantly outperforms other baseline methods across all models and datasets.
These gains might be attributed to Wikipedia’s inherent tree-like hierarchical structure, which enables \myname to holistically model document-level dependencies as metadata.

\subsection{Ablation Study} 
We evaluate the effectiveness of contextual signals and structured metadata through ablation studies by removing the corresponding representations from the retrieval formulation in Eq~\ref{eq:retrieval}.
Ablation results are visualized in Fig~\ref{fig:abalation}, from which we have several findings.
First, the complete \myname framework (with both contextual signals and structured metadata) consistently outperforms its variant using only the representations of knowledge chunks themselves.
This demonstrates that explicitly modeling document-level, multi-granular context and metadata strengthens retrieval-side semantics, particularly enhancing recall capability through complementary information fusion.
Second, the relative importance of these components varies across domains: contextual signals contribute more to performance gain on SciFact, while document-level metadata is more useful for TrevCOVID and NFCorpus datasets.
The results of our ablation study confirm that \myname’s multi-channel encoding effectively leverages both latent contextual patterns and explicit structural knowledge.

\subsection{Top-$k$ Retrieval Analysis}

Fig.~\ref{fig:topk} presents the experimental results under varying Top-$k$ retrieval settings, from which we draw the following observations. 
First, the left panel of Fig.~\ref{fig:topk} demonstrates the performance trajectories of different models and methods on the same dataset as $k$ increases. We evaluate several commonly used $k$ values in RAG systems (1, 3, 5, 10, 15). The results reveal that compared to naive RAG, our \myname maintains consistent performance improvements across all $k$ values. Furthermore, naive RAG exhibits noticeable performance degradation with larger $k$ values, likely due to excessive redundant information in retrieved content.
Second, the right panel of Fig.~\ref{fig:topk} illustrates the performance variation of \myname across different datasets. Notably, our method demonstrates positive correlation between larger $k$ values and improved answer F1 scores on most datasets. 
These experimental results indicate that \myname effectively balances comprehensive retrieval with generation efficiency and accuracy, successfully mitigating the common performance deterioration issue observed in baseline methods when processing larger retrieval sets.

\section{Conclusion}
In this paper, we identify a limitation in existing RAG methods: the use of identical knowledge chunk representations for both retrieval and generation, despite their distinct requirements. 
To address this, we propose \myname, a heterogeneous RAG framework that decouples knowledge representations to optimize retrieval accuracy as well as generation efficiency and efficacy simultaneously. 
By leveraging multi-granular contextual signals and metadata for retrieval and concise chunks for generation, our approach mitigates redundancy while preserving critical knowledge. 
Furthermore, we propose an adaptive prompt-tuning strategy for the retrieval model to adapt the heterogeneous retrieval augmented generation process.
Extensive experiments across retrieval tasks and end-to-end generation pipelines validate that \myname significantly outperforms baseline methods. 
These results highlight the importance of tailoring knowledge representations to the unique demands of retrieval and generation steps. 
In general, this work provides a principled direction for advancing RAG systems by harmonizing the dual objectives of retrieval precision and generation quality.

\clearpage

\section*{Limitations}
While \myname demonstrates promising results, this work has two main limitations that suggest directions for future research.
First, the experimental validation currently focuses on several widely-used benchmark datasets from selected domains. 
Although these datasets represent important application areas for RAG systems, our findings may not fully generalize to emerging domains with distinct knowledge characteristics. 
Future work should validate the of \myname across more diverse domains and emerging application contexts.
Second, our framework primarily focuses on optimizing the retrieval side knowledge chunk representations, while employing relatively straightforward represention for generation side. 
Prompt token compression techniques could potentially better preserve critical information while further improving generation efficiency. 
This presents a promising direction for subsequent research to enhance the generation-side optimization while maintaining the decoupling paradigm of our framework.
We exclusively utilize generative AI to refine the writing and verify grammatical accuracy in this paper.



\bibliography{custom}




\end{document}